**Dimensionality Control of *d*-orbital Occupation in Oxide Superlattices**

Da Woon Jeong[1,2], Woo Seok Choi[3,4], Satoshi Okamoto[3], Jae –Young Kim[5], Kyung Wan Kim[6], Soon Jae Moon[7], Deok–Yong Cho[1,2,8], Ho Nyung Lee[3] & Tae Won Noh[1,2]

[1]Center for Correlated Electron Systems, Institute for Basic Science (IBS), Seoul 151-747, Korea

[2]Department of Physics and Astronomy, Seoul National University, Seoul 151-747, Korea

[3]Materials Science and Technology Division, Oak Ridge National Laboratory, Oak Ridge, Tennessee 37831, United States

[4]Department of Physics, Sungkyunkwan University, Suwon, Gyeonggi-do 440-746, Korea

[5]Pohang Accelerator Laboratory, Pohang University of Science and Technology, Pohang 790-784, Korea

[6]Department of Physics, Chungbuk National University, Cheongju 361-763, Korea

[7]Department of Physics, Hanyang University, Seoul 133-791, Korea

[8]Department of Physics, Chonbuk National University, Jeonju 561–756, Korea

**Manipulating the orbital state in a strongly correlated electron system is of fundamental and technological importance for exploring and developing novel electronic phases. Here, we report an unambiguous demonstration of orbital occupancy control between $t_{2g}$ and $e_g$ multiplets in quasi-two-dimensional transition metal oxide superlattices (SLs) composed of a Mott insulator $LaCoO_3$ and a band insulator $LaAlO_3$. As the $LaCoO_3$ sublayer thickness approaches its fundamental limit (i.e. one unit-cell-thick), the electronic state of the SLs changed from a Mott insulator, in which both $t_{2g}$ and $e_g$ orbitals are partially filled, to a band insulator by completely filling (emptying) the $t_{2g}$ ($e_g$) orbitals. We found the reduction of dimensionality has a profound effect on the electronic structure evolution, which is, whereas, insensitive to the epitaxial strain. The remarkable orbital controllability shown here offers a promising pathway for novel applications such as catalysis and photovoltaics, where the energy of *d* level is an essential parameter.**

Transition metal oxides (TMOs) offer a considerable number of emergent physical phenomena, such as superconductivity, magnetism, and (multi)ferroicity. These phenomena originate from the complex interplay among charge, spin, orbital, and lattice degrees of freedom associated with the $d$ states of transition metals[1-3]. Due to their inherently coupled nature, even a small change in the orbital degree of freedom can bring about drastic evolution of the underlying electronic structure[2-5]. Therefore, finding a tuning knob for controlling the relative occupancy of $d$ orbitals has been a central issue to tailor the intriguing physical properties of TMOs. The outcome of such a study, thus, not only deepens our fundamental understandings but also allows us to envision novel TMO-based electronic devices[3-5].

Recent advances in atomic-scale synthesis of TMO heterostructures and artificial superlattices (SL) offer unprecedented opportunities for the control of orbital states that cannot be realized within bulk counterparts. Examples include (1) controlling the physical interaction and relevant orbital energy levels at the interface between a cuprate and a manganite, resulting in a systematic tuning of the orbital occupation[5, 6]; (2) inducing high-$T_c$ cuprate-like orbital states in a LaNiO$_3$/LaAlO$_3$ perovskite heterostructure by controlling the electron occupation of particular $d$ states with the $e_g$ symmetry ($x^2$-$y^2$ and $3z^2$-$r^2$)[7-9]; and (3) modifying the $e_g$ orbitals systematically to optimize the catalytic activities at the surface of TMOs[10-12]. The charge transfer needed for a catalytic reaction was found to be facilitated by the $d$-level near the chemical potential.

Up to date, however, the orbital control has been rather limited to a subset of $d$ electronic states, i.e. either within $t_{2g}$ or $e_g$ states. Despite a huge number of reports on the control of the electron number within the respective $t_{2g}$ or $e_g$ states (e.g. Refs. [7,9]), the artificial control of relative electron populations across the two subsets has not been realized because the $t_{2g}$-$e_g$ separation, called 10$Dq$, is usually a few eV (in typical perovskite TMOs), much larger than the energy scale we can engineer e.g., with heterostructuring. In this Report, we demonstrate that a dramatic control over the occupation between the $t_{2g}$ and $e_g$ orbital states can be achieved in quasi-2-dimensional (2D) artificial SLs of a TMO heterostructure. The SLs are composed of $LaCoO_3$ (LCO) as an active layer and $LaAlO_3$ (LAO) as a spacer layer. LCO is a unique perovskite which has a "twin" of quantum many-body states. In the case of bulk LCO, a low spin (LS) state ($t_{2g}^6$; $S = 0$) is the ground state and a high spin (HS) state ($t_{2g}^4 e_g^2$; $S = 2$) is the first excited state, with an energy difference less than 50 meV[13-15] (see Supplementary Information for details of the discussions on the intermediate spin state[16]). At room temperature, bulk LCO has a mixed spin state of 40% HS + 60% LS due to thermal excitation[13-15]; namely, ~1 electron (hole) occupies the $e_g$ ($t_{2g}$) orbital in bulk LCO. Therefore, the room temperature electronic state of bulk LCO has been considered as a Mott insulator[14, 17]. The energies of the twin states are in delicate balance and thus the electron populations of those states could be tuned with a minimal energy cost by applying strain or controlling dimensionality through heterostructuring.

**Results**

Figure 1 shows a schematic diagram of our approach to control the inter-multiplet orbital occupation in LCO. At the ultrathin limit, one of the $e_g$ orbital levels, namely, $3z^2-r^2$ ($z$ is the direction of the surface normal, i.e. along the crystallographic [001] direction) can be modulated to change the insulating nature of LCO. In particular, the quantum confinement effect can selectively increase the energy of orbitals concentrated along the $z$ direction ($3z^2-r^2$ and $xz/yz$). Moreover, in heterostructures with a spacer layer such as LAO, certain chemical interactions at the interface between the LCO layer and the spacer layer can influence the strength of the Co $3d(3z^2-r^2)$-O $2p$ hybridization[3, 8, 18]. Such interactions can affect the total energy of the HS state while keeping the total energy of the LS state almost unchanged. Thus, heterostructuring as in a quasi-2D oxide system offers a means to modify the delicate population balance between two quantum many-body states (HS and LS).

This scenario in ultrathin LCO is clearly supported by the results of the dynamical mean field theory (DMFT) calculations[19]. Figure 2 comparatively shows the electronic structures of an ideal bulk (3-dimension; 3D) and a single layer (2D) LCO system (for details of the calculations, see Supplementary Information[16]). The total/orbital-resolved density of states (DOS) for the 3D and 2D systems are presented in Figures 2a and 2b, respectively. The corresponding momentum-resolved spectral functions are shown in Figures 2c and 2d. The energy distribution of the orbital states shows broad Co $e_g$ (narrow Co $t_{2g}$) bands located between +1 and +4 eV (-1 and +1 eV). O $2p$ bands are mostly located below -2 eV. In the case of 3D LCO, a portion of the $e_g$ orbital is occupied (the red-shaded area in the DOS near $E$ = -0.8 eV), while a portion of the $t_{2g}$ orbital is unoccupied (the blue-shaded area in DOS just above the Fermi level). This configuration manifests a mixed spin state in 3D LCO at room temperature[14, 15]. In contrast, in the 2D case, it is clearly shown that the $e_g$ bands are empty and the $t_{2g}$ bands are almost fully occupied, suggesting an electron transfer from the $e_g$ orbital to the $t_{2g}$ orbital.

It should be noted that the total DOS near $E$ = -0.5 eV is significantly enhanced as the dimensionality is reduced (3D → 2D; from Figure 2a to Figure 2b). This is primarily due to the increase in the number of oxygen ions that participate in the O $p$ – Co $d$ hybridization in 2D LCO; in 3D LCO, the total DOS comprises 3(O $p_\sigma$) + 6(O $p_\pi$) + 2(Co $e_g$) + 3(Co $t_{2g}$), while in 2D LCO, 4(O $p_\sigma$) + 8(O $p_\pi$) + $(x^2-y^2)$ + $(3z^2-r^2)$ + $(xy)$ + 2$(yz/zx)$. Therefore, the charge transfer nature becomes much enhanced in 2D LCO even with a small O $p_\pi$ DOS at $E$ = -0.5 eV.

The redistribution of the *d*-orbital occupation in dimension-controlled LCO yields a dramatic change in the nature of the electronic ground state from a Mott insulator to a band insulator. In the spectral function of the 3D LCO sample (Figure 2c), nearly non-dispersive and incoherent states (denoted by a yellow arrow) are observed just above the Fermi energy. These states have the Co $t_{2g}$ orbital character, reflecting that the 3D LCO is a Mott insulator, as we denoted the optical transition across the Mott gap by the arrow *a*. In contrast, the 2D LCO sample is found to be a band insulator, in which the gap at the Fermi level is defined by the energy separation between the fully occupied $t_{2g}$ and empty $e_g$ bands (instead of $t_{2g}$); the corresponding optical transition is represented by the arrow $\beta$. It is also shown that the lower $e_g$ band edge along the Γ-X direction shifts to higher energy in the 2D LCO case. The blueshift of the transition $\beta$ is presumably caused by the fact that the coordination of the $Co^{3+}$ ion with the upper and lower layers is absent, and the hopping in the out-of-plane direction is suppressed. The dimensional control can stimulate the electronic quantum phase transition as well as the sizable redistribution of the orbital occupation in the 2D LCO sample.

In order to experimentally validate the theoretical prediction on the control of orbital occupation, we designed TMO SLs composed of ultrathin LCO layers embedded within the large band-gap insulator LAO. Systematic dimensional control was conducted by reducing the periodicity of SLs below a few atomic unit-cells; namely, $(LCO)_n/(LAO)_n$ SLs (*n* = 2, 6, and 10). The LCO SLs were fabricated on $NdGaO_3$ (NGO) (110) substrates using pulsed laser epitaxy (see Supplementary Information[16]). Figure 3 shows x-ray diffraction reciprocal space maps of the three samples (*n* = 2, 6, and 10) near the (103) reflection of the substrate. The indices on the satellite peaks reflect the periodicity of the SLs. For more detailed structural analyses, see Supplementary Information[16].

It is clearly shown that all of the H values in the HKL-coordinate of the SL peaks (red spots) were unity, indicating that the *in-plane* lattice constants were coherently maintained with respect to the substrate. This result suggests negligible influence of strain among the LCO SLs. Note that the lattice mismatch of bulk LCO (pseudocubic) and LAO on a NGO substrate is 1.6% and 1.9%, respectively. However, it has also been shown that the electronic structure in LCO thin films is rather insensitive to the lattice strain, while the magnetic properties showed strong dependence[20]. The insensitiveness against the strain might be due to the peculiar electronic structure of LCO. As shown in Figure 2, there exists O 2$p$ density of states near the Fermi level. According to our DMFT study on the strain dependence (not shown), the strain mostly influences the O 2$p$ energies but not much the Co 3$d$ energies. This suggests that valence, and consequently, ionic size of Co$^{3+}$ can hardly change under external stresses. Therefore, we can tell unambiguously that the evolution in the electronic structure found in this work, is originated mainly from the dimensionality rather than the strain effects.

The evolution in electronic structure predicted by the DMFT calculations is verified experimentally using optical spectroscopy. Figure 4 shows optical conductivity ($\sigma(\omega)$) of the three SLs. For comparison, $\sigma(\omega)$ of a 13 nm-thick LCO film is attached. We observed strong absorption features at $\hbar\omega$ ~3 eV (thin gray lines) and 1.5 eV (thin yellow triangles), together with a weak absorption feature at $\hbar\omega$ ~ 0.5 eV (filled red triangles). Based on the results of the DMFT calculations (Figure 2), we can attribute these features to a O $p$-Co $d$ charge transfer transition ($\gamma$ ~ 3 eV), a Co $d(t_{2g})$-$d(e_g)$ transition ($\beta$ ~ 1.5 eV), and a Co $d(t_{2g})$-$d(t_{2g})$ transition ($\alpha$ ~ 0.5 eV), respectively. It is clearly observed that as $n$ decreases, the weak absorption at $\hbar\omega$ ~0.5 eV ($\alpha$) is systematically suppressed and disappears at (LCO)$_2$/(LAO)$_2$. Based on the DMFT results, the disappearance of the excitation $\alpha$ can be explained in terms of the crossover of the spin-orbital ground state from a mixed HS + LS configuration to a LS configuration with decreasing the dimension, which is generically accompanied by a crossover from a Mott insulator to a band insulator; the optical transition between the Hubbard bands of Co $t_{2g}$ states disappears with decreasing $n$, and the optical gap is defined by the transition from fully occupied Co $t_{2g}$ bands to empty $e_g$ bands. Also, the blueshift of transition $\beta$, predicted by the DMFT calculation results, is indeed observed in the experimental $\sigma(\omega)$, as denoted by the yellow triangles in Figure 4. The excellent agreement between theoretical and experimental results confirms that our SLs indeed undergo the dimensional crossover of the electronic structure, as illustrated in Figure 1.

**Discussion**

Polarization-dependent O $K$-edge X-ray absorption spectroscopy (XAS) data further reveal the details of orbital occupation change in our (LCO)$_n$/(LAO)$_n$ SLs with variation in the dimensionality (see the Supplementary Information for details[16]). Since the O $K$-edge XAS probes the transition to the unoccupied O 2$p$–Co 3$d$ hybridized states, we can monitor the change in the 3$d$ orbital occupancy of Co$^{3+}$ ions directly. Figure 5a shows the isotropic XAS spectra ([($E$ // $c$) + 2 × ($E$ // $ab$)]/3) measured at room temperature. The spectrum from a 13 nm-thick LCO film (gray dots) is included in the figure for comparison. The main features near the photon energy $\hbar\omega$ = 529 eV and the shoulder near $\hbar\omega$ = 528 eV are attributed to the unoccupied Co $e_g$ and Co $t_{2g}$ states, respectively[15,21,22]. Features near $\hbar\omega$ = 532 eV reflect the Al 3$sp$ states in the LAO substrate. In contrast to the case of thick LCO, the unoccupied Co $t_{2g}$ state nearly disappears with decreasing $n$, while the main structure of the unoccupied Co $e_g$ state is clearly enhanced as shown by the arrow in Figure 5a. This spectral evolution provides concrete evidence of the electron transfer from the $e_g$ state to the $t_{2g}$ state.

Moreover, the polarization dependence in the XAS data reveals that the $e_g$-to-$t_{2g}$ electron transfer can be facilitated by an increase in the $3z^2$-$r^2$ orbital energy. Figure 5b highlights the anisotropy in the electronic structure. The polarization-dependent XAS spectra for $n = 2$ and $n = 10$ SLs are shown in Figure 5b. The ($E // c$) data are shown as dots while the ($E // ab$) data are shown as solid lines. The $e_g$ peak in the ($E // c$) spectra are located at slightly higher energies compared with their respective peak in ($E // ab$) counterparts, suggesting that the $3z^2$-$r^2$ state is at a higher level than the $x^2$-$y^2$ state.

The corresponding X-ray linear dichroism [XLD = ($E // c$)-($E // ab$)] is displayed in Figure 5c with a dotted zero-line for guidance to eyes. The spectral difference is more clearly observed in the XLD spectra. The positive (negative) signs in the XLD indicate the dominance of the $3z^2$-$r^2$ ($x^2$-$y^2$) character. The XLD signal is enhanced with decreasing $n$ from 10 to 2. Such anisotropy in the $e_g$ orbitals manifests 'lifting' of the $3z^2$-$r^2$ level under low dimensionality[5,8,12,23,24]. The energy difference between the $3z^2$-$r^2$ state and the $x^2$-$y^2$ state is estimated to be ~1 eV, which is very close to the energy distance between the leading peak positions in the unoccupied $3z^2$-$r^2$ and $x^2$-$y^2$ DOS in the DMFT result (see Figure 2b). For more clear assignments, we also performed Co $L_{2,3}$-edge XAS and the corresponding XLD measurements. The analysis of the Co $L_{2,3}$-edge data using cluster model calculations confirmed the XLD results of the changes in the $e_g$ orbital states (see Supplementary Information for more details[16]).

Our combined theoretical and experimental results clearly revealed how the $d$-orbital occupation with different symmetries ($t_{2g}$ and $e_g$) can be manipulated by two-dimensional layering of atomically thin Mott insulator-band insulator SLs. By controlling the delicate competition between the crystal field splitting and electron correlation effect, we found that a subtle change in the $3z^2$-$r^2$ orbital could be induced as a consequence of a $d$-orbital reconstruction from a $t_{2g}^{6-\delta}e_g^{\delta}$ state to a complete $t_{2g}^{6}$ state. The lattice relaxation for all of the LCO SLs was not observed, implying that the evolution in the electronic structure is indeed driven by the reduced dimensionality rather than the lattice strain effect. The reduced dimensionality altered the insulating nature of LCO with a spin state crossover from LS+HS to LS. Therefore, our demonstrated controllability of electron occupation between the $t_{2g}$ and $e_g$ orbital states with atomic-scale SLs manifests that an accurate growth control of oxide heterostructures can open up many opportunities for discovering new physical properties or functionalities such as catalysis and photovoltaics, where the energy of $d$ level is an essential parameter.

**Methods**

*Sample fabrication*. $(LCO)_n/(LAO)_n$ SLs ($n$ = 2, 6, and 10) were grown on $NdGaO_3$ (110) substrates by pulsed laser epitaxy at 700 °C under an oxygen atmosphere (100 mTorr). A KrF excimer laser ($\lambda$ = 248 nm) with a laser fluence of ~1 J cm$^{-2}$ was used to ablate sintered LCO and single-crystalline LAO targets. The crystallinity of the fabricated SL samples were confirmed by lab x-ray diffraction (XRD) and hard XRD at the BL14B1 beamline of the Shanghai Synchrotron Radiation Facility in China.

*Optical spectroscopy*. Optical conductivity spectra ($\sigma(\omega)$) were obtained by reflectance measurements in the energy region of 0.1–1.0 eV and variable angle spectroscopic ellipsometry (JA Woollam V-VASE) in the energy region of 0.74-4.5 eV. The $\sigma(\omega)$ of the LCO layers in the SLs were obtained by optical simulations implemented in the W-VASE software. The optical constants of bulk LAO were measured independently for the SL optical model calculations.

*X-ray absorption spectroscopy*. Soft X-ray absorption spectroscopy (XAS) measurements were performed at the 2A beamline at the Pohang Light Source (PLS) in Korea, which is equipped with an elliptically polarized undulator (EPU). The energy resolution was approximately 0.1 eV. We measured the absorption coefficients in both the fluorescence yield (FY) and total electron yield (TEY) modes. Note that we have not observed any discernable difference between the TEY and FY data in the energy region of interest in the O $K$-edge XAS spectra. Here, we changed the polarization of light by tuning the phase and gap in the EPU while fixing the angle of incidence with respect to the surface normal to the SLs to 70°, rather than by rotating the samples. This process circumvents any issues associated with the finite probing depth of the soft X-rays. The spectra of $E // \sigma$ reflect only in-plane orbital hybridization, whereas the spectra of $E // \pi$ reflect mostly (88%) perpendicular hybridization and partially (12%) in-plane orbital hybridization. The isotropic term ($[(E // c) + 2\times (E // ab)]/3$) and the XLD (($E // c$) - ($E // ab$)) at the Co $L_{2,3}$-edge were deduced from the raw data of ($E //\sigma$) and ($E //\pi$) using the relationships of ($E //\sigma$) = ($E // ab$) and ($E //\pi$) = $\sin^2 70°$ ($E // $ c)+ $\cos^2 70°$ ($E // ab$), respectively. The spectra were normalized maintaining the total spectral weight after subtracting the contributions of the $NdGaO_3$ substrates.

**Acknowledgments**

We thank G. A. Sawatzky, B. Keimer, V. Hinkov, S. S. A. Seo, S. H. Chang, and R. Eder for helpful discussions. This work was supported by the Institute for Basic Science (IBS) in Korea (x-ray and optical spectroscopy) and by the U.S. Department of Energy, Basic Energy Sciences, Materials Sciences and Engineering Division (sample design by pulsed laser epitaxy and theory).

**Author contribution statement**

D.W.J. and W.S.C. conceived and designed the experiments, and D.-Y.C., H.N.L. and T.W.N. supervised the research. W.S.C and H.N.L. fabricated the superlattice samples, and D.W.J. and D.-Y.C. performed the spectroscopic measurements. S.O. performed the DMFT calculations. All authors contributed to the analyses of the experimental data and the writing of the manuscript.

**Additional information.**

**Figure Captions**

**Figure 1** Reducing dimensionality to control the *d*-orbital population in $(LaCoO_3)_n(LaAlO_3)_n$. Two competing spin-orbital configurations, i.e. low-spin (S = 0; $t_{2g}^6$) and high-spin (S = 2; $t_{2g}^4 e_g^2$), are shown for the $Co^{3+}$ ions. The two spin states in 3D bulk $LaCoO_3$ are nearly degenerate, resulting in a mixed-spin state at room temperature, whereas the low-spin state can be induced in quasi-2D superlattices by changing the energies of the $3z^2$-$r^2$ and *xz/yz* states.

**Figure 2** Dynamical mean field theory results. Total/orbital-resolved density of states (DOS) and spectral functions as a function of energy and momentum for (a, c) 3D and (b, d) 2D configurations, respectively. Each orbital-resolved DOS is normalized so that the total area becomes unity. A yellow arrow in (c) denotes the nearly dispersionless $t_{2g}$ band. Possible optical transitions *α*, *β*, and *γ* are indicated by arrows.

**Figure 3** X-ray diffraction reciprocal space maps of the three $(LaCoO_3)_n(LaAlO_3)_n$ superlattices with *n* = 2, 6, and 10, near the (103) reflection of the substrate. The indices on the satellite peaks reflect the superlattice periodicity. All of the *H* values in the *HKL*-coordinate of the superlattice peaks (red spots) were unity, suggesting negligible influence of strains on the atomic structure of the $LaCoO_3$ superlattices.

**Figure 4** Optical conductivity spectra of $(LaCoO_3)_n(LaAlO_3)_n$ superlattices (*n* = 2, 6, and 10) and a 13 nm-thick $LaCoO_3$ film. With decreasing *n*, the optical transition within Co $t_{2g}$ (i.e., $t_{2g} \rightarrow t_{2g}$) was suppressed, opening a charge gap. Thin lines are fitting curves highlighting each peak.

**Figure 5** (a) O *K*-edge XAS isotropic spectra of $(LaCoO_3)_n(LaAlO_3)_n$ superlattices (*n* = 2, 6, and 10) and the thick $LaCoO_3$ film taken in a fluorescence yield (FY) mode. The intensity of the $t_{2g}$ ($e_g$) feature near 528 eV (529 eV indicated by a vertical dotted line) decreases (increases) with decreasing *n*, suggesting a dramatic transition to a low-spin state. The measurement geometry is appended in the inset. (b) Polarization-dependent spectra and (c) the X-ray linear dichroism (XLD) spectra highlighting the orbital character. Features near 532 eV reflect the Al 3*sp* states in the $LaAlO_3$ substrates.

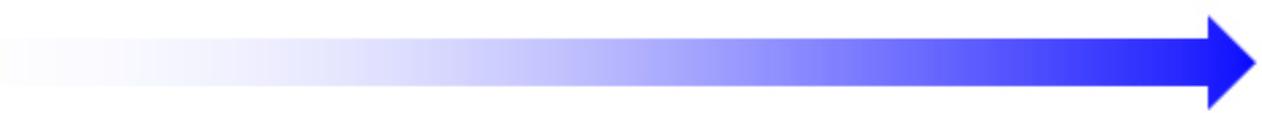

$n = \infty$            $n = 2$

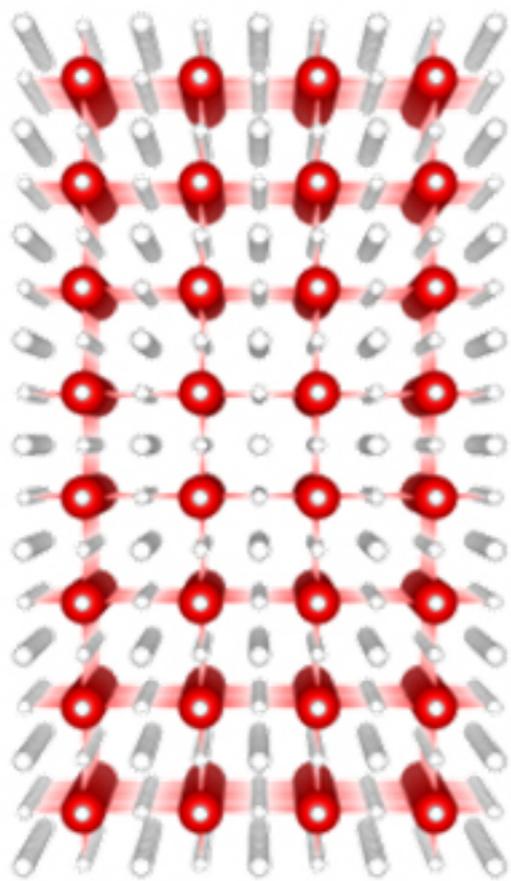

LaCoO$_3$

*3D*

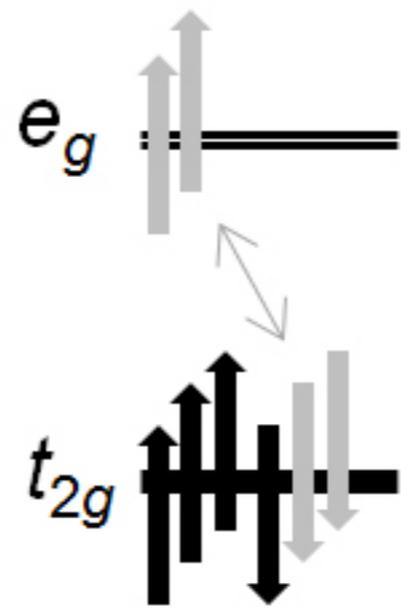

High Spin + Low Spin    Low Spin

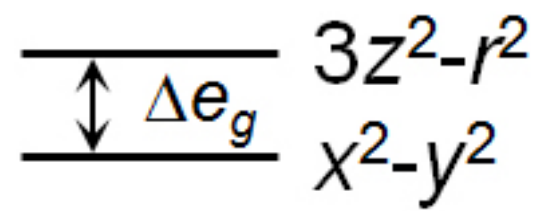

$e_g$ — $3z^2-r^2$, $x^2-y^2$, $\Delta e_g$

$t_{2g}$ — $xz, yz$, $xy$

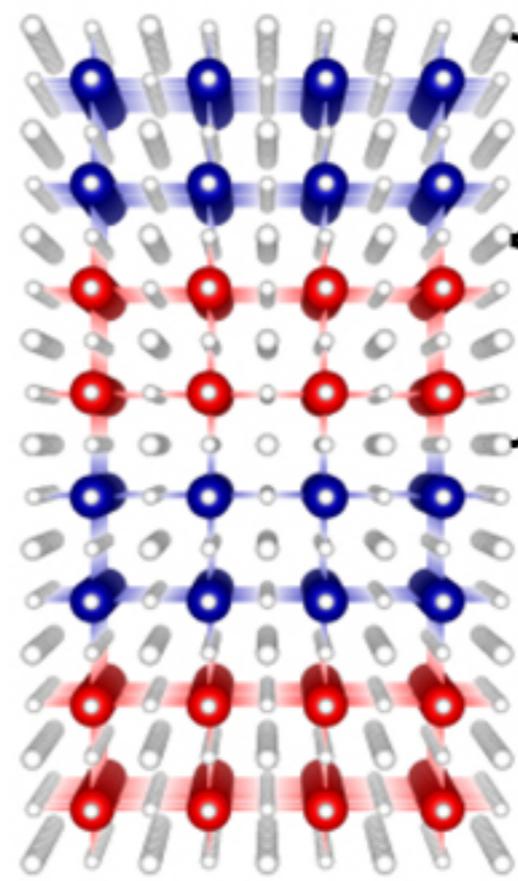

LaAlO$_3$

LaCoO$_3$

*quasi-2D*

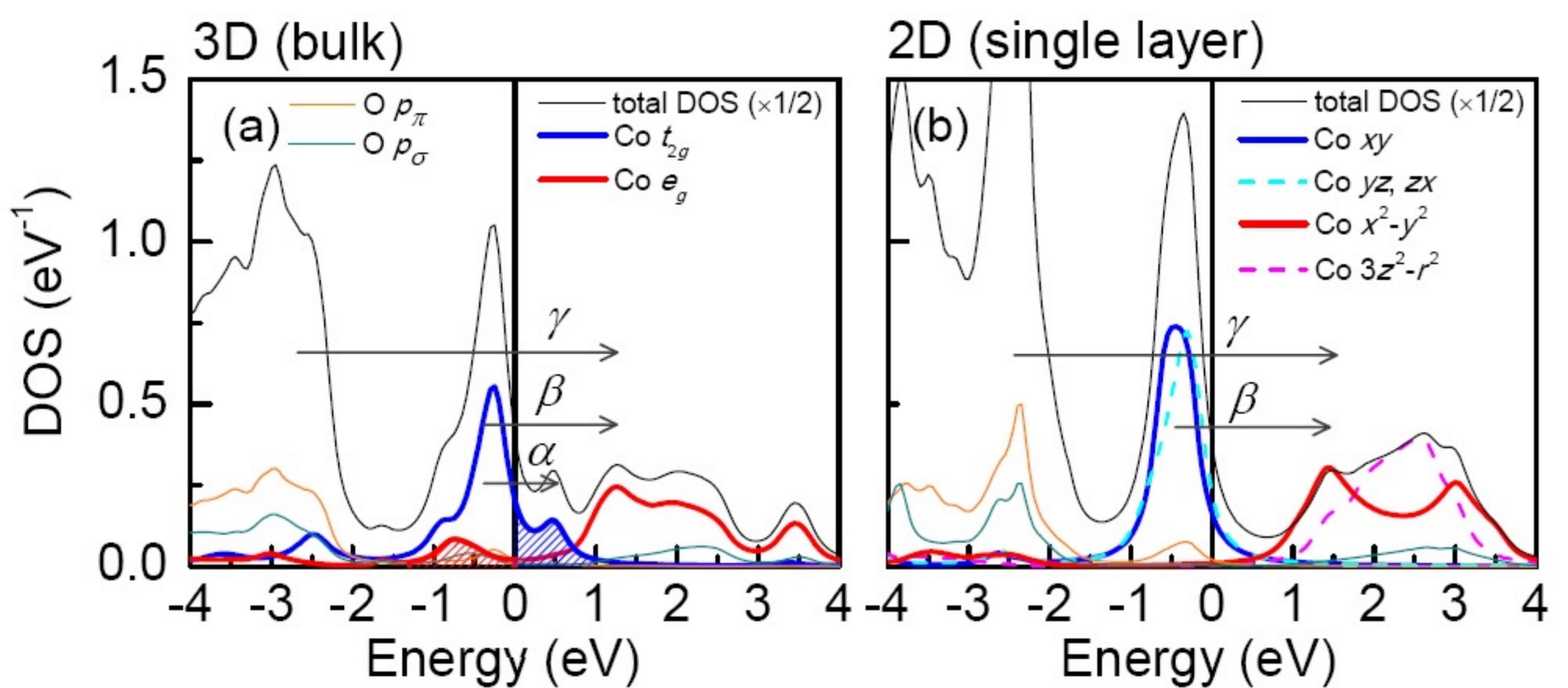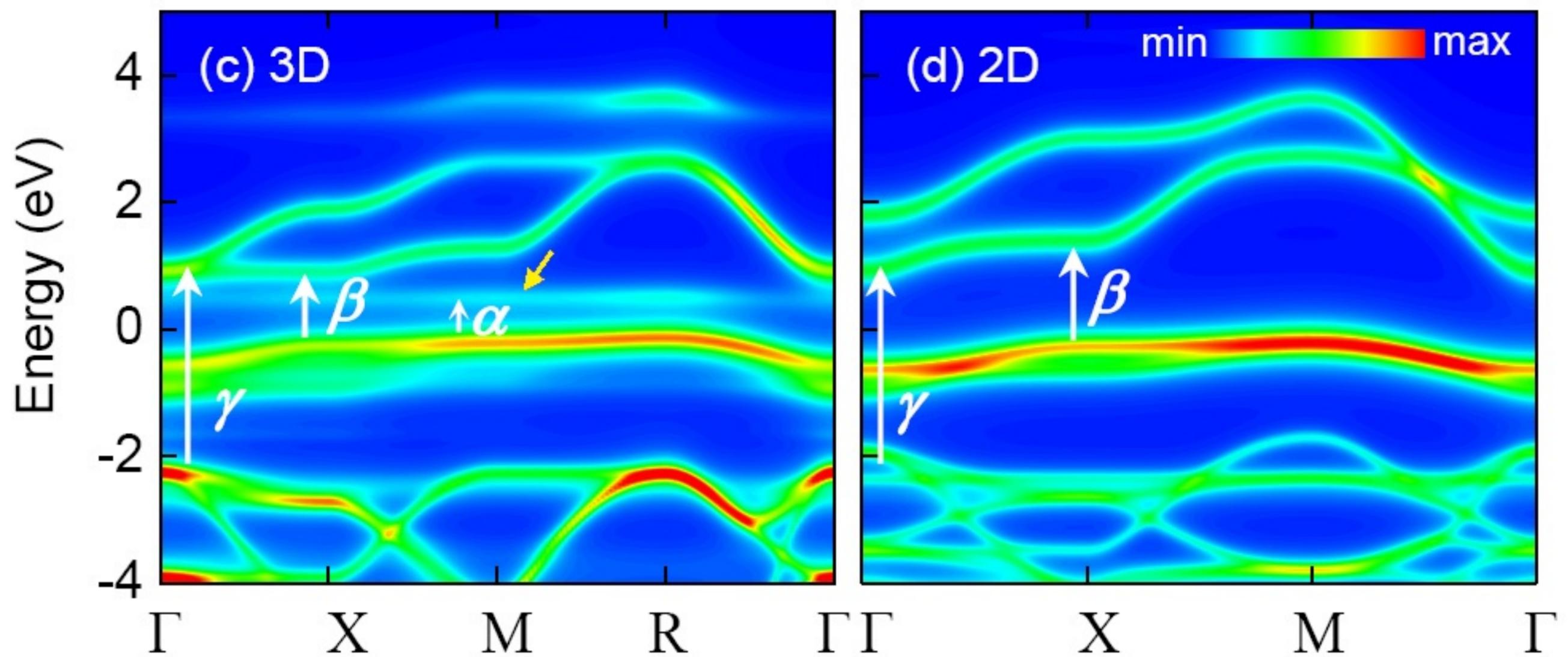

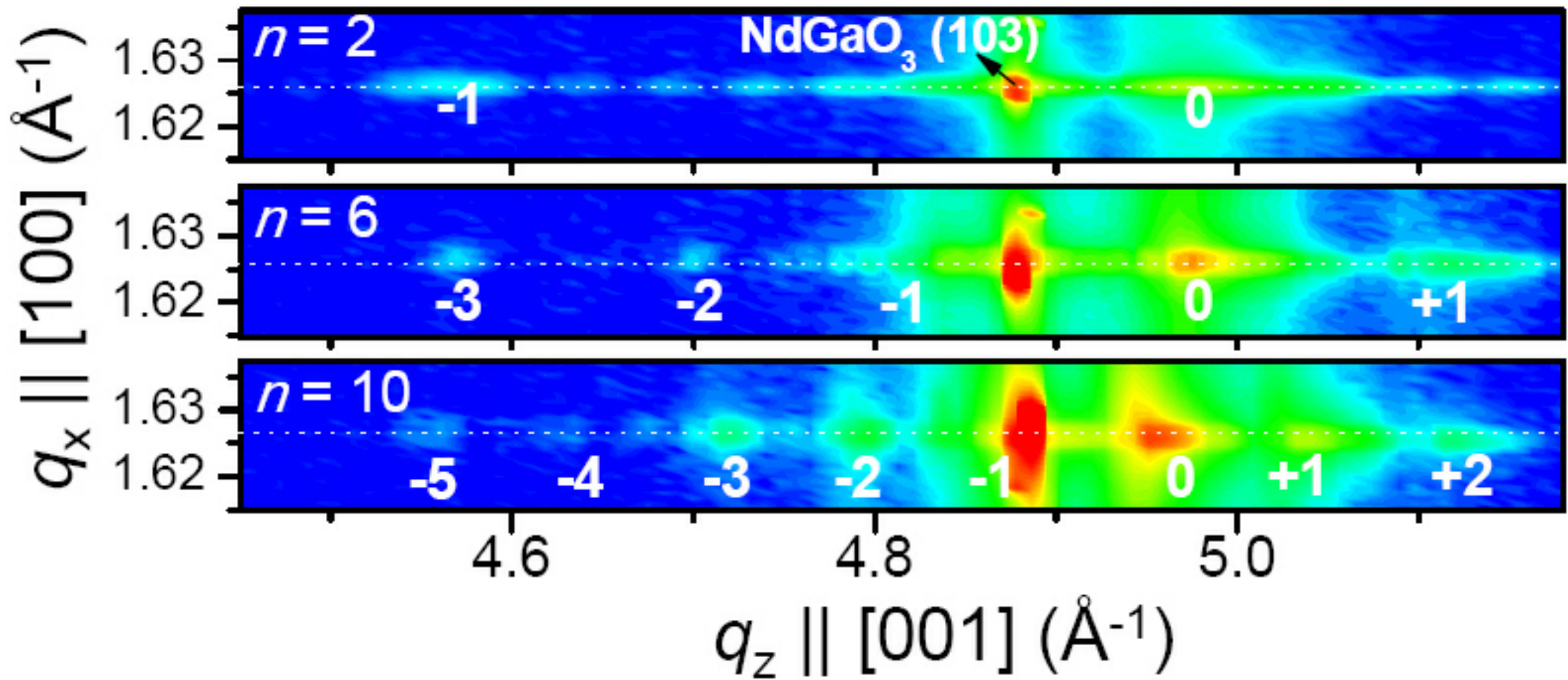

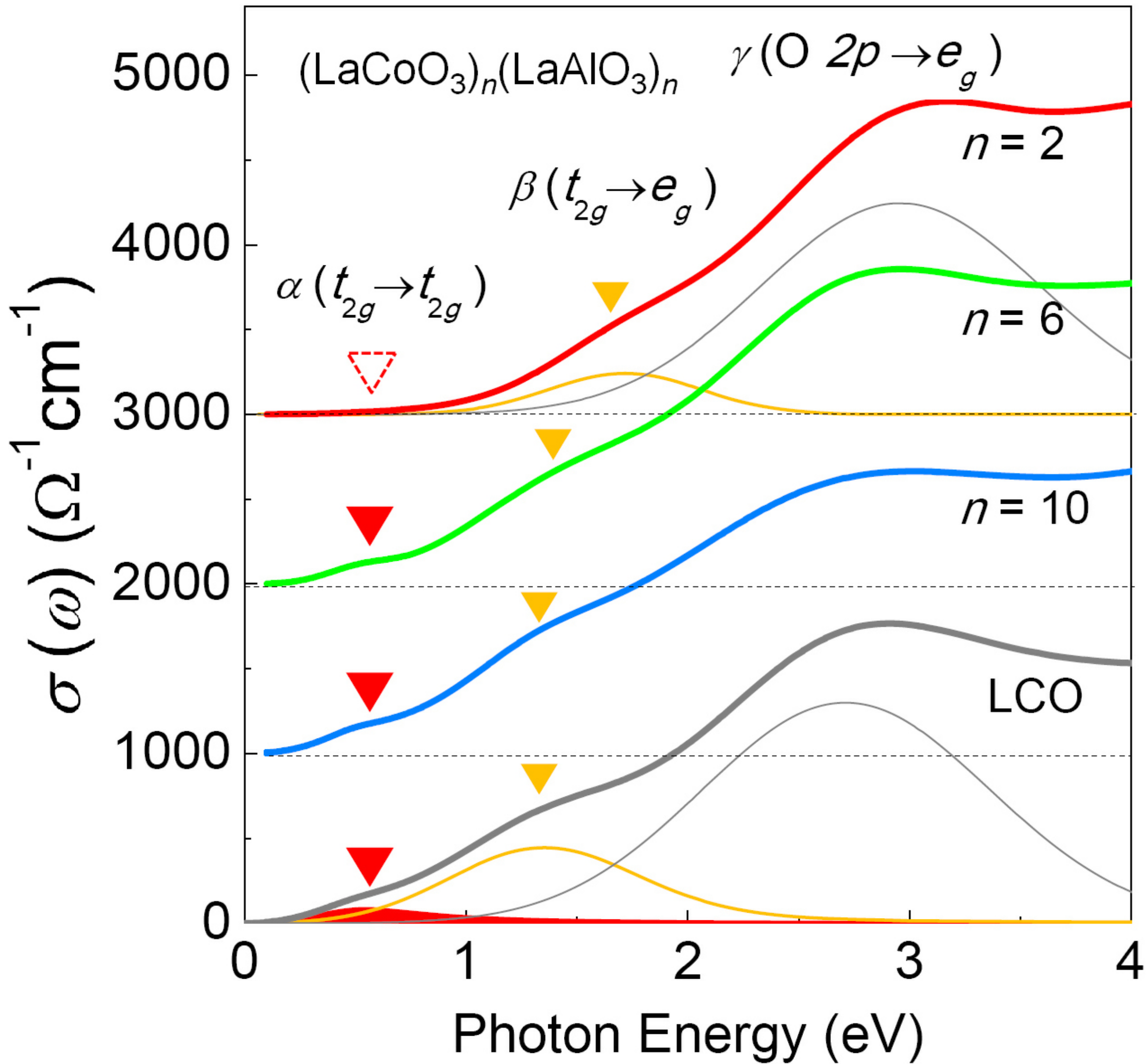

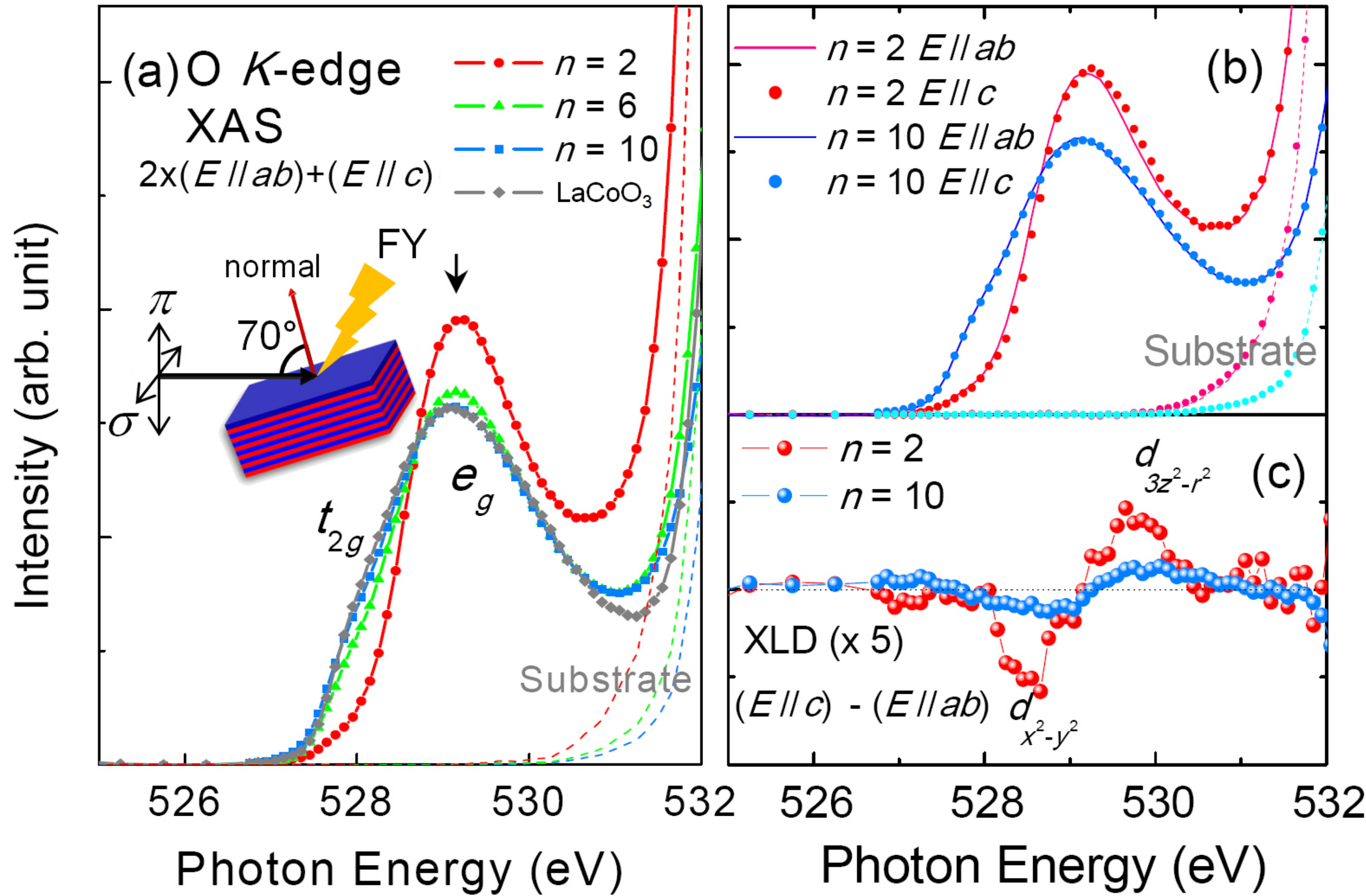

Supplementary Information

# Dimensionality Control of *d*-orbital Occupation in Oxide Superlattices

D. W. Jeong[1,2], W. S. Choi[3,4], S. Okamoto[3], J.–Y. Kim[5], K. W. Kim[6], S. J. Moon[7], Deok–Yong Cho[1,2,8*], H. N. Lee[3†], and T. W. Noh[1,2]

[1]*Center for Correlated Electron Systems, Institute for Basic Science (IBS), Seoul 151-747, Korea*

[2]*Department of Physics and Astronomy, Seoul National University, Seoul 151-747, Korea*

[3]*Materials Science and Technology Division, Oak Ridge National Laboratory, Oak Ridge, Tennessee 37831, United States*

[4]*Department of Physics, Sungkyunkwan University, Suwon, Gyeonggi-do 440-746, Korea*

[5]*Pohang Accelerator Laboratory, Pohang University of Science and Technology, Pohang 790-784, Korea*

[6]*Department of Physics, Chungbuk National University, Cheongju 361-763, Korea*

[7]*Department of Physics, Hanyang University, Seoul 133-791, Korea*

[8]*Department of Physics, Chonbuk National University, Jeonju 561–756, Korea*

Electronic Addresses: D.-Y. Cho (zax@jbnu.ac.kr), H. N. Lee (hnlee@ornl.gov)

I. Methods

**Sample fabrication.** $(LCO)_n/(LAO)_n$ SLs ($n$ = 2, 6, and 10) were grown on $NdGaO_3$ (110) substrates by pulsed laser epitaxy at 700 °C under an oxygen atmosphere (100 mTorr). A KrF excimer laser ($\lambda$ = 248 nm) with a laser fluence of ~1 J cm$^{-2}$ was used to ablate sintered LCO and single-crystalline LAO targets. The crystallinity of the fabricated SL samples were confirmed by lab x-ray diffraction (XRD) and hard XRD at the BL14B1 beamline of the Shanghai Synchrotron Radiation Facility in China.

**Optical measurement.** Optical conductivity spectra ($\sigma(\omega)$) were obtained by reflectance measurements in the energy region of 0.1–1.0 eV and variable angle spectroscopic ellipsometry (JA Woollam V-VASE) in the energy region of 0.74–4.5 eV. The $\sigma(\omega)$ of the LCO layers in the SLs were obtained by optical simulations implemented in the W-VASE software. The optical constants of bulk LAO were measured independently for the SL optical model calculations.

**X-ray absorption spectroscopy.** Soft X-ray absorption spectroscopy (XAS) measurements were performed at the 2A beamline at the Pohang Light Source (PLS) in Korea, which is equipped with an elliptically polarized undulator (EPU). The energy resolution was approximately 0.1 eV. We measured the absorption coefficients in both the fluorescence yield (FY) and total electron yield (TEY) modes. Note that we have not observed any discernable difference between the TEY and FY data in the energy region of interest in the O $K$-edge XAS spectra. Here, we changed the polarization of light by tuning the phase and gap in the EPU while fixing the angle of incidence with respect to the surface normal to the SLs to 70°, rather than by rotating the samples. This process circumvents any issues associated with the finite probing depth of the soft X-rays. The spectra of $E // \sigma$ reflect only in-plane orbital hybridization, whereas the spectra of $E // \pi$ reflect mostly (88%) perpendicular hybridization and partially (12%) in-plane orbital hybridization. The isotropic term ([ (*E* // *c*) + 2× (*E* // *ab*)]/3) and the XLD ((*E* // *c*) - (*E* // *ab*)) at the Co $L_{2,3}$-edge were deduced from the raw data of (*E* // $\sigma$) and (*E* // $\pi$) using the relationships of (*E* // $\sigma$) = (*E* // *ab*) and (*E* // $\pi$) = $\sin^2 70°$ (*E* // *c*) + $\cos^2 70°$ (*E* // *ab*), respectively. The spectra were normalized maintaining the total spectral weight after subtracting the contributions of the $NdGaO_3$ substrates.

**DMFT calculations.** The electronic properties of LCO are analyzed using dynamical mean field theory (DMFT) [1], which maps an interacting bulk problem into a quantum impurity problem to be solved self-consistently. For the single-electron part of the Hamiltonian, we use a multi-orbital tight-binding (TB) model based on the linear combination of atomic orbitals method. The TB model consists of Co 3*d* orbitals and O 2*p* orbitals, with band parameters derived in Ref.[2]: (*pd$\sigma$*) = 1.776, (*pd$\pi$*) = −0.975, (*pp$\sigma$*) = 0.93, (*pp$\pi$*) = −0.112, $\varepsilon_d$* (bare *d*-level energy) = −46.4, and $\varepsilon_d$ (Hartree-shifted *d*-level) = 1.731 (eV). We determined the appropriate 10*Dq* value of 1.1 eV, which locates the bulk 3D system in the mixed spin-state between the HS and LS configurations[3]. We considered an ideal perovskite structure, which has periodic boundary conditions (PBCs) along the *x*, *y*, and *z* directions for the 3D system. A finite thick slab was considered for the 2D system, in which a Co site was coupled with two apical O sites located above and below; open-boundary conditions were considered along the *z* direction, as well as two equatorial O sites with the PBCs. A rotationally symmetric Coulomb interaction is considered for the Co 3*d* electrons described in terms of Racah parameters[2, 3]: *A* = 7.62, *B* = 0.14, and *C* = 0.54 (eV). The impurity problem consisted of the interacting Co *d* orbitals coupled with the effective medium. In this study, the effective medium was approximated as a finite number of bath sites with one bath per correlated orbital, and the impurity model was solved using the Arnoldi algorithm[4-6]. Here, we considered only paramagnetic solutions with temperature $k_BT$ = 0.02 eV, to retain low-energy states with Boltzmann factors larger than $10^{-6}$. We note that the finite DOS at Fermi energy of 3D system is due to the finite broadening 0.1eV of the self-energy in DMFT with exact diagonalization impurity solver.

## II. STRUCTURAL PROPERTIES

We fabricated $(LCO)_n/(LAO)_n$ SLs on $NdGaO_3$ (NGO) (110) substrates using pulsed laser epitaxy. The X-ray diffraction (XRD) reciprocal space maps for the three samples ($n = 2, 6,$ and 10) near (103) in the ($H, K, L$) pseudocubic coordinates of NGO are shown in Fig. 3 in the text and are also reproduced in Fig. S1. The peaks with a null index (denoted in the figure) originated from the averaged lattice constants of LCO and LAO while those with nonzero indices correspond to the satellite peaks produced by repetition of the $(LCO)_n(LAO)_n$ sublattices. The small peaks between the satellite peaks were Kiessing fringes, which suggest the high quality of the SLs. All of the $H$ values of the peaks were unity, indicating that the *in-plane* lattice constants were coherently maintained with respect to the substrate. The tensile strain on both LCO and LAO due to the NGO substrate was estimated to be $\varepsilon \approx$ 1.6% and 1.9%, respectively. The effect of this strain on the electronic structure of LCO was studied previously; however, no noticeable changes were observed [7]. The overall XRD results indicate that the designed SL structure had been successfully obtained and that these SLs were of high quality. The Bragg peaks become broader as $n$ increases, reflecting the increasing roughness or inhomogeneity of the SLs.

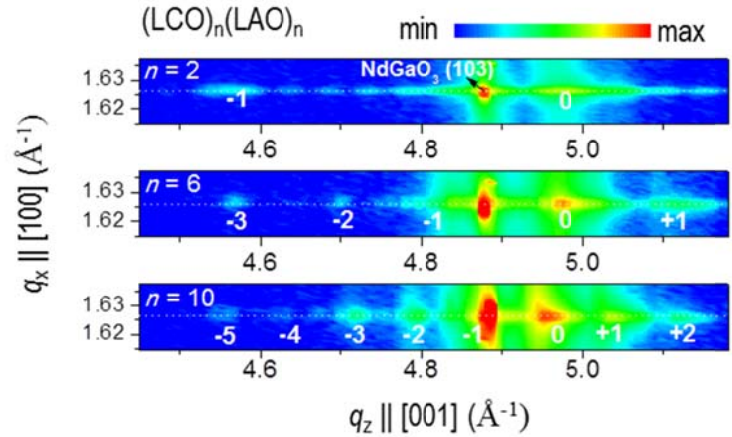

Figure S1 reproduces Fig. 3 in the text.

## III. DMFT CALCULATION

Figure S2a presents the average orbital occupancies of the three-dimensional (3D) and two-dimensional (2D) configurations, with several values of 10$Dq$ indicated. Lowering the dimensionality induces an electron transfer from the $e_g$ state to the $t_{2g}$ state; the behaviour is rather robust for those values of 10$Dq$ ~ 1 eV. The small deviation in the orbital occupancy could originate from the quantum confinement effect or hybridization effect between the Co 3$d$ and the O 2$p$ states. The change in the spin state was also confirmed in the square mean value of the total spin moment $\langle S_z^2 \rangle$ (Fig. S2b). More importantly, the DMFT correctly predicted the dimensional crossover of the spin state due to the change in orbital occupancy. Whereas the 2D system showed the same 10$Dq$ dependence as the 3D system for both orbital occupancy and spin moment, the LS state was more favourable for the 2D system and the HS-LS transition took place at a smaller 10$Dq$ (~0.2 eV smaller).

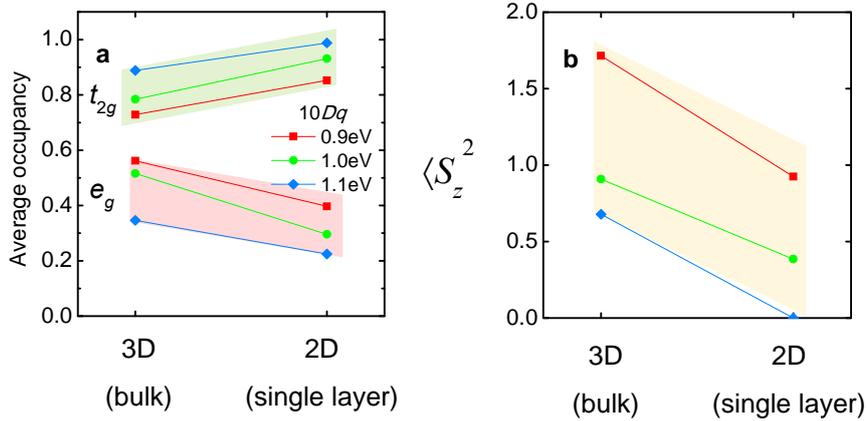

Figure S2 Dynamical mean field theory (DMFT) calculation results with various 10$Dq$'s. (a) Average orbital occupancy and (b) spin moment for the 3-D and 2-D systems with various 10$Dq$ indicated. The shades are guides to the eye.

## IV. SPIN STATE AT HIGH TEMPERATURE

$Co^{3+}$ ion in LCO has six *d*-electrons which can in principle have spin-orbital configurations of high spin (HS; $t_{2g}^4 e_g^2$: S = 2), low spin (LS; $t_{2g}^6$: S = 0) and intermediate spin (IS; $t_{2g}^5 e_g^1$: S = 1) states. The ground state (zero temperature) spin-orbital configuration is LS. However, at high temperature, the spin configuration becomes nontrivial due to inclusion of HS or IS.

Historically, Racah and Goodenough[8, 9] first argued that the high temperature evolution of spin configuration should originate from inclusion of the HS state. The total energy difference between HS and LS configuration is much small (less than 80meV)[8]. The temperature evolution of crystalline structure, magnetic susceptibility, and transport properties was successfully interpreted by this so-called LS-HS scenario[8]. On the other hand, Korotin *et al.*[10] argued based on the results of LDA+U calculation that IS state can also be stabilized, so that the first excited state could be IS rather than HS (LS-IS scenario).

After the long standing controversy (see V. Krapek *et al.*,[11] and R. Eder *et al.*,[2] for more details), it is now broadly accepted that the LS-HS scenario is much more favorable with lots of supporting experimental data including XMCD[12], neutron scattering[13] and electron-spin-resonance. Recent theoretical studies with variational cluster approximation and dynamical mean field theory also show consistency with the LS-HS scenario. The IS is found to have much higher energy than HS so that the IS contribution can be neglected. The excellent agreement between our DMFT prediction and experimental results can also be a new supporting data for the LS-HS scenario.

## V. Co $L_{2,3}$ EDGE XAS AND CLUSTER MODEL CALCULATION

To obtain more detailed information on the orbital occupations and spin states of the SLs, we performed Co $L_{2,3}$-edge XAS measurement at room temperature in total electron yield. The probing depth of the total electron yield is large enough to probe the buried interfaces of $(LCO)_n(LAO)_n$ SLs because there is no absorption at Co $L_{2,3}$ edges from the LAO layers. The isotropic term ($[(E // c) + 2\times (E // ab)]/3$) and the XLD (($E // c$) - ($E // ab$)) at the Co $L_{2,3}$-edge were deduced from the raw data of ($E // \pi$) and ($E // \sigma$) using the relations of ($E // \sigma$) = ($E // ab$) and ($E // \pi$) = $\sin^2 70^\circ$ ($E // c$) + $\cos^2 70^\circ$ ($E // ab$), respectively.

The Co $L_3$ and $L_2$ white lines are located near $\hbar\omega$ = 780 and 790 eV, respectively. In the isotropic spectra, the peak energies do not change significantly with $n$ (Fig. S3a), suggesting that the chemical valence of the Co $d$ states are nearly identical ($Co^{3+}$) for all SLs[12, 14, 15]. However, details in the features are different from each other; the overall features are sharpened and the spectral weights of the satellite peaks at 792 eV are reduced with decreasing $n$. Haverkort *et al.* reported a similar evolution in the Co $L_3$ and $L_2$ XAS spectra of bulk LCO with decreasing temperature[12]. They interpreted the spectral evolution in terms of spin-state crossover from a mixed-spin state of HS and LS into a pure LS. The $n$-dependent spectral changes shown in Fig. S3a are similar to that of the earlier temperature-dependent studies, suggesting that the spin state in the LCO SL thin films can indeed be controlled by periodicity ($n$). Figure S3b shows the Co $L_{2,3}$-edge XLD spectra of the SLs. The XLD spectra, which highlights the anisotropy in the orbital configuration, show an appreciable evolution in the peak feature near $\hbar\omega$ = 779 eV and the dip near $\hbar\omega$ = 781 eV with $n$.

The spectral evolutions in the XLD spectra were analyzed using a cluster model calculation, which includes the configuration interactions and the full atomic multiplets. The cluster model calculation of the Co $L_{2,3}$-edge XAS and XLD were performed using CTM4XAS software[16] to fit the experimental spectra. First, the best-fit condition for the $n = 10$ sample was obtained with the parameters of $10Dq = 1.41$ eV, $Dt = 0$, $Ds = -0.01$ (for bare crystal fields), $pd\sigma = 1.776$ eV, $pd\pi = 0.975$ eV (for the transfer integrals), $\Delta = 2$eV (for the charge transfer energy), and $U_{dd} = 5.5$ eV and $U_{pd} = 7.0$ eV (for the on-site Coulomb interactions). The values of the hybridization strengths are adopted from the values employed in the DMFT calculation. Meanwhile, the $10Dq$ value is about 0.3 eV larger than used in the DMFT calculation. The larger value is plausibly originated from neglecting the crystal fields for the oxygen sites for simplicity.

For perfect cubic perovskite, the orbital configuration should be isotropic; thus, the XLD should be zero. Thus, the XLD in the thick SL ($n = 10$) should originate from tetragonal distortion due to the tensile strain in the LCO layers. For the low-$n$ samples (i.e., $n = 2$ and 6), the strength of the O $2p$-Co $3z^2$-$r^2$ hybridization (originally 1.776 eV) was increased additionally just to obtain a good fit to the experimental XLD spectra. We increased the hybridization strength $pd\sigma$ from 1.776 eV to 1.79 eV and 1.81 eV to fit the XLD/XAS spectra of $n = 6$ and $n = 2$, respectively. Such an increase in the hybridisation would increase the energy difference between $3z^2$-$r^2$ and $x^2$-$y^2$ for the $n = 2$ ($n = 6$) sample further, in addition to the energy splitting from the tensile strain[17]. The stronger hybridization of O $2p$-Co $3z^2$-$r^2$ compared with O $2p$-Co $x^2$-$y^2$ can be explained by considering the average strength of the metal–oxygen hopping integrals within the SLs. In the LCO sub-layer, the $3d$ orbitals of the Co$^{3+}$ ions are strongly hybridized by the adjacent O $2p$ orbitals; thus, the Co$^{3+}$-O$^{2-}$ covalent bonds are rather strong. In contrast, in the LAO sub-layer, the Al$^{3+}$-O$^{2-}$ ionic bonds are much weaker. Thus, the oxygen ions in the Co$^{3+}$-O$^{2-}$-Al$^{3+}$ bond chain at the interface would move slightly toward the Co$^{3+}$ ions[18]. Although this ionic movement may be small, it can substantially increase the hybridization strength of the Co $d$-O $p$ perpendicular to the SL plane ($xz/yz$ and $3z^2$-$r^2$) compared to the in-planar Co $d$-O $p$ hybridization.

Meanwhile, the quantum confinement itself (due to loss of periodicity of Co ions along $z$-axis) can also lift up the $3z^2$-$r^2$ orbital without involving the chemical interaction at the interface of LCO/LAO. Practically, we cannot distinguish its effect from the chemical influence (hybridization effect). Therefore, we should understand the apparent increase of hybridisation used in the cluster model calculation only as parameter control to mimic the increase in energy of $3z^2$-$r^2$ orbital.

It should be noted that the actual energy difference between $3z^2$-$r^2$ and $x^2$-$y^2$ observed in O K-edge XAS is approximately 1eV. We observe in the DOS of 2D LCO (Fig. 2b in the text) that the $x^2$-$y^2$ DOS is split into two features while the $3z^2$-$r^2$ DOS is not split and lie in-between. The increment of hybridization parameter [($pd\sigma$//$z$) - ($pd\sigma$//$x,y$)] used in the Co $L_{2,3}$-edge XAS simulation (~0.1eV) refers to the difference in the mean energy values rather than peak-to-peak difference. The increase in the energy of $3z^2$-$r^2$ orbital (~1eV) is large enough to break the delicate balance between the Hund exchange energy and 10$Dq$. Therefore, the thermal population of the HS states at room temperature became strongly suppressed. Our XLD data demonstrate that the drastic modification to a complete LS state can be achieved by judiciously manipulating the competing many-body ground states of LS and HS.

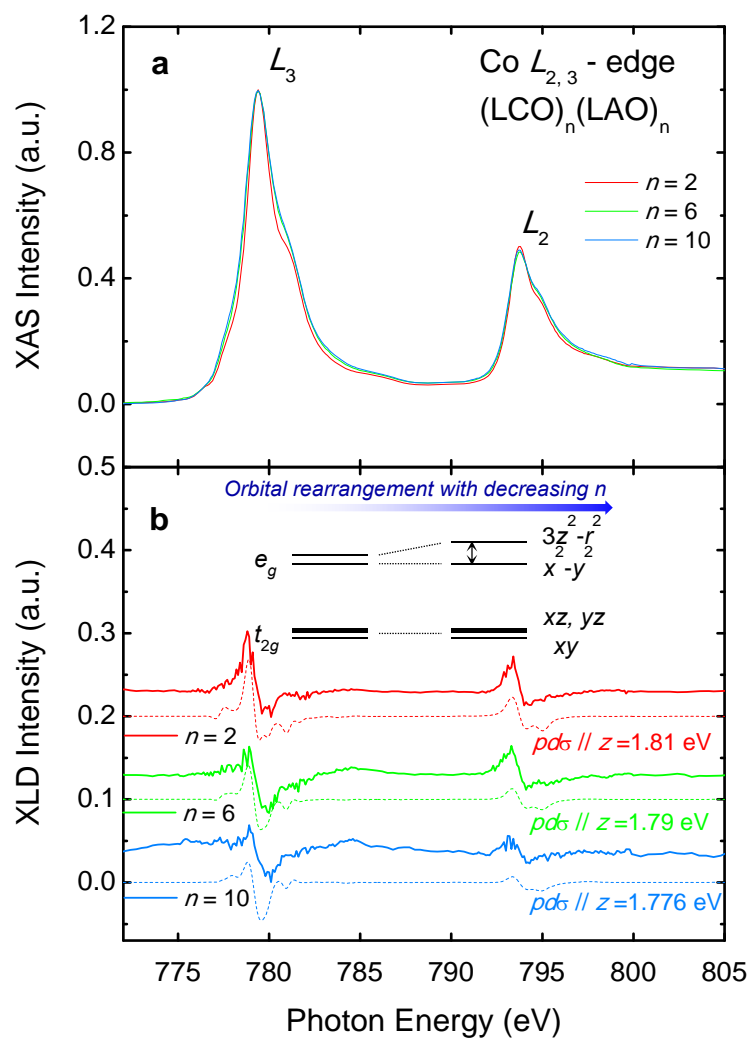

Figure S3 Co $L_{2,3}$ edge XAS and XLD spectra and cluster model calculation results. (a) Isotropic spectra $[(E // c) + 2\times(E // ab)]/3$ and (b) XLD $[(E // c) - (E // ab)]$ measured in the total electron yield mode. In the simulation, we increased the hybridization strength of O $2p$- Co $3d$ ($3z^2$-$r^2$) ($pd\sigma$//$z$) from 1.776 eV to 1.81 eV only to mimic the increase of $3z^2$-$r^2$ energy. The simulation results are in a good agreement with the experimental spectra.

## VI. MAGNETIZATION MEASUREMENT

To check the magnetic properties of the LCO/LAO SLs, we have fabricated series of LCO/LAO SLs on $(LaAlO_3)_{0.29}(SrAl_{0.5}Ta_{0.5}O_3)_{0.71}$ (LSAT) substrates which show almost same lattice parameters with NGO. The purpose of using the LSAT substrate is to circumvent the strong paramagnetic signals from the Nd ions in the NGO substrate. The films on the LSAT substrates were not good for the spectroscopic measurements, because the substrate characteristics were slightly modified after we deposited the films by PLD method, especially at high photon energy. This made it difficult to subtract the substrate signal after we obtain the optical conductivity spectra of thin films. After the deposition, the optical absorption near 3.5 eV varied possibly due to the oxygen vacancy formation in the LSAT substrate.

Figure S4a and S4b present the temperature and magnetic field dependence of the magnetization of the samples, respectively. With decreasing temperature, we observed ferromagnetic ordering of Co spins below $T_c$ ~80K which is consistent with the previous studies on epitaxial LCO films.[7] The magnetic ordering is significantly suppressed for 2D LCO samples. Probably, it is due to the suppression of HS population in 2D LCO samples as we have proved using spectroscopic measurements. When the number of Co ions increases three times and five times (with increasing $n$ from 2 to 6 and 10), the measured magnetic signal increases significantly. This implies the increase of HS population with increasing layer thickness.

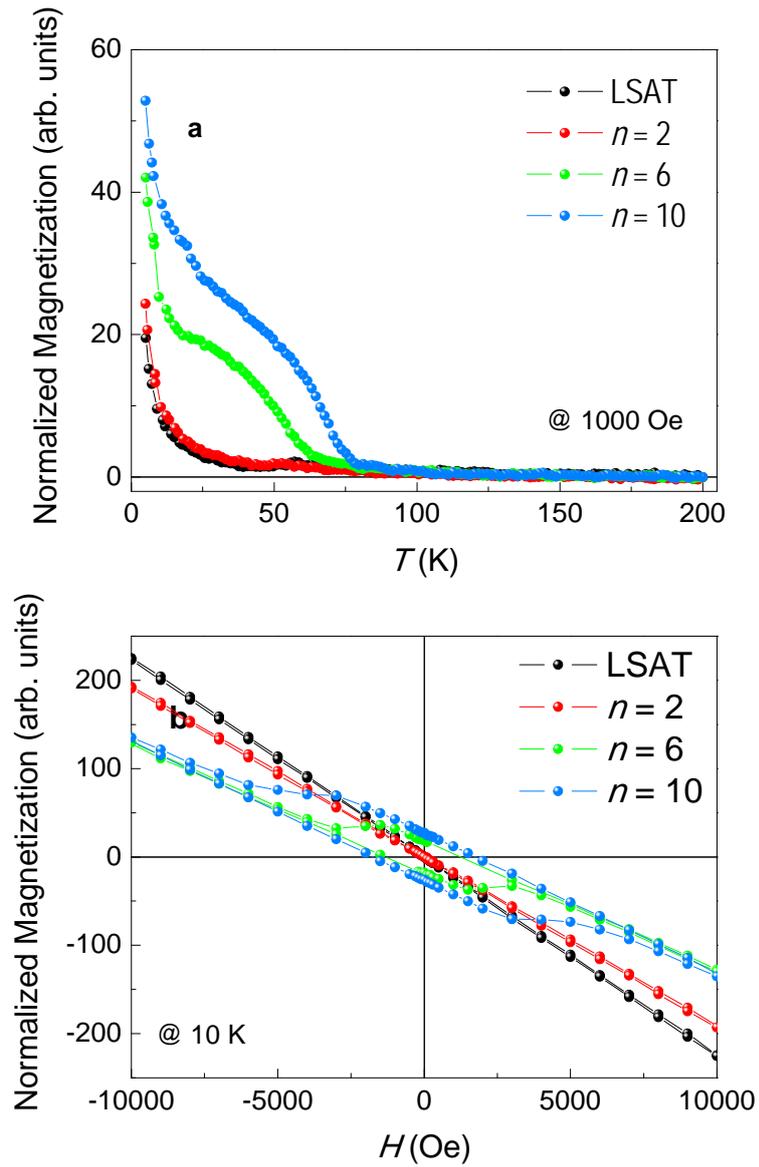

Figure S4 (a) temperature and (b) magnetic field dependent magnetization of the LCO SLs on LSAT **substrates.** The signals from the ferromagnetic order become weaker significantly as *n* decreases, implying the suppression of HS population in 2D LCO samples.